%% file: main.tex
\documentclass[twoside]{ilcws08}
\usepackage[latin1]{inputenc}
\usepackage[dvips]{graphicx,epsfig,color}
\usepackage{wrapfig,rotating}
\usepackage{amssymb,amsmath,array}
\usepackage{booktabs}
\usepackage{subfigure}

\def\zh2nnh{$ZH \to \nu\bar{\nu}H$}
\def\zh2qqh{$ZH \to q\bar{q}H$}
\def\zh2qqcc{$ZH \to q\bar{q}c\bar{c}$}

\renewcommand{\thesubsection}{\thesection.\alph{subsection}}

\newcommand{\be}{\begin{equation}}
\newcommand{\ee}{\end{equation}}
\newcommand{\bea}{\begin{eqnarray}}
\newcommand{\eea}{\end{eqnarray}}
\newcommand{\gsim}{\hbox{ \raise3pt\hbox to 0pt{$>$}\raise-3pt\hbox{$\sim$} }}
\newcommand{\lsim}{\hbox{ \raise3pt\hbox to 0pt{$<$}\raise-3pt\hbox{$\sim$} }}
\newcommand{\mathbold}[1]{\mbox{\boldmath $\bf#1$}}

\def\to{\rightarrow}

\def\gsim{\,\lower.25ex\hbox{$\scriptstyle\sim$}\kern-1.30ex%
\raise 0.55ex\hbox{$\scriptstyle >$}\,}
\def\lsim{\,\lower.25ex\hbox{$\scriptstyle\sim$}\kern-1.30ex%
\raise 0.55ex\hbox{$\scriptstyle <$}\,}

\begin{document}
\input yonamine

\newpage
\end{document}

%% file: yonamine.tex
\setcounter{section}{0}
\setcounter{figure}{0}
\setcounter{table}{0}
\setcounter{footnote}{0}

\begin{center}
    {\large\bf A study of top-quark Yukawa coupling measurement\\
      in $\mathbold{\it e^+e^-\rightarrow t \bar{t}H$ at $\sqrt{s} = {\rm 500}\,{\rm GeV}}$}\\
\vspace{0.8cm}
        {\bf Ryo Yonamine}$^{(a)}$,
        {\bf Katsumasa Ikematsu}$^{(b)}$,
        {\bf Satoru Uozumi}$^{(c)}$, and
        {\bf Keisuke Fujii}$^{(b)}$

    \vspace{0.8cm}

    {\it
     $^{(a)}${The Graduate University for Advanced Studies, Tsukuba, Japan} \\
     $^{(b)}${IPNS, KEK, Tsukuba, Japan} \\
     $^{(c)}${Department of Physics, Kobe University, Kobe, Japan}
   }
    \vspace{0.8cm}

\begin{quote} \small
We report on the feasibility of measuring the top Yukawa coupling in the process: $e^+e^-\rightarrow t\bar{t}H$. This measurement is crucial to test the mass generation mechanism for matter particles. Since the cross section for this process attains its maximum around  $\sqrt{s}=700\,$GeV, most of the past studies were done assuming this energy region. It has been pointed out, however, that the QCD threshold correction enhances the cross section significantly and might enable its measurement at  $\sqrt{s}=500\,$GeV, which will be accessible already in the first phase of the ILC project. We have implemented this threshold enhancement into our $t\bar{t}H$ event generator and carried out Monte Carlo simulations. Our results show that $t\bar{t}H$ events can be observed with a significance of $4.1\,\sigma$ with no beam polarization and $5.4\,\sigma$ with the $e^-$ and $e^+$ beam polarization combination: $(-0.8,+0.3)$. 
\end{quote}

  \end{center}.

\section{Introduction}
The standard model of elementary particle physics is based on two pillars: one is the gauge principle and the other is the electroweak symmetry breaking and  mass generation mechanism. 
The first pillar, the gauge principle, has been tested by precision electroweak measurements. 
On the other hand, the second pillar has not yet been tested. 
In order to confirm this second pillar we have to measure the Higgs self-coupling and the top Yukawa coupling. 

In this study, we investigate the feasibility of measuring the top Yukawa coupling at 500$\,$GeV with the process: $e^+e^-\rightarrow t\bar{t}H$. 
Since the top quark is the heaviest among all the matter particles, the measurement of its Yukawa coupling will be the most decisive test of the mass generation 
mechanism for matter particles.
 Since the cross section for the $e^+e^-\rightarrow t\bar{t}H$ process is 2-3$\,$fb even near its maximum reached at around  $\sqrt{s}=700\,$GeV, most of the past studies assumed the measurement energy in this region (\cite{Ref:paststudy1},\cite{Ref:paststudy2}). 
 It has been pointed out, however, that the QCD threshold correction enhances the cross section significantly and might open up the possibility of  measuring the top Yukawa coupling at  $\sqrt{s}=500\,$GeV, which is within the scope of the first phase of the ILC project(\cite{Ref:QCDcorr1}-\cite{Ref:QCDcorr17}).
 In order to investigate this possibility we have implemented this threshold enhancement into our $t\bar{t}H$ event generator and carried out Monte Carlo simulations. 
 
 In the next section we begin with clarifying the signatures of the $t\bar{t}H$ production and list up possible background processes that might mimic the signal.
 We then describe our analysis framework used for event generations and detector simulations in section 3. 
 The event selection procedure for the generated events is elaborated in section 4, considering characteristic features of the background processes.
 The results of the event selection are given in section 5.
 Section 6 summarizes our results and concludes this report. 
 
\section{Signal and Possible Background}
\label{sec:signature}
\begin{figure}[h]
          \begin{center}
            \includegraphics[scale=0.8,clip]{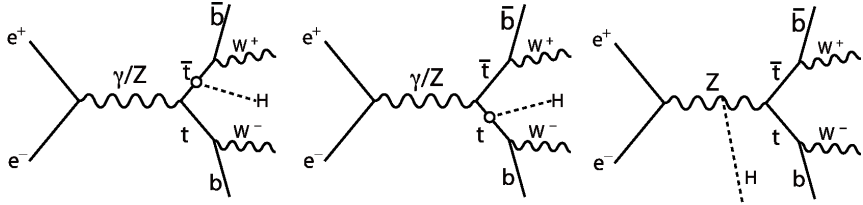}
           \end{center}
           \caption{
              Feynmann diagrams for the $t\bar{t}H$ process
           }
           \label{fig:TTHdiagram}
   \end{figure}

The Feynman diagrams for the $e^+e^-\to t\bar{t}H$ process followed by $t(\bar{t})\to b(\bar{b})W$ decays are shown in Figure \ref{fig:TTHdiagram}.
Notice that the first and second diagrams contain the top Yukawa coupling, which we want to measure. 
The signatures of $t\bar{t}H$ events depend on how the $H$ and the $W$s decay.
In this study we concentrate on the dominant decay mode: $H\rightarrow b\bar{b}\, (68\%)$.
The signal events hence have four $b$ jets and two $W$s. 
The $t\bar{t}H$ events can then be classified into 3 groups (8-jet, 1-lepton+6-jet, and 2-lepton+4-jet modes) corresponding to the combinations of leptonic and hadronic  decays of the two $W$s.
For $W$s that decayed leptonically we cannot reconstruct their invariant masses due to missing neutrinos. 
On the other hand, for the $W$s that decayed hadronically we can reconstruct their masses and use them as a signature.
For the $t$ or the $\bar{t}$ with a hadronically decayed $W$ we can also use the invariant mass of the 3-jet system to test if it is consistent with the top mass.

\begin{figure}[h]
        \begin{center}
          \includegraphics[scale=0.5,clip]{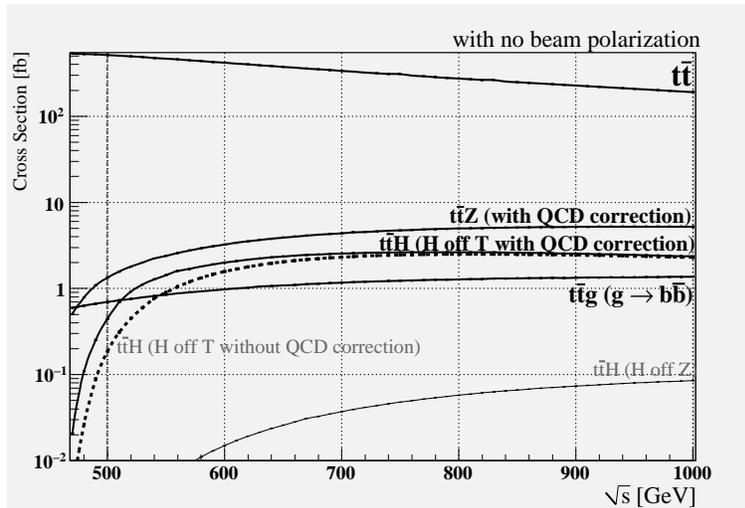}
          \caption{
             Production cross section of the signal, $t\bar{t}H$, together with those of the main background processes, 
             $t \bar{t}H,  t\bar{t}Z, t\bar{t}, t\bar{t}g$, as a function of the center of mass energy for no beam polarization.
          }
          \label{fig:xsection}
        \end{center}  
      \end{figure}

Possible background processes that might mimic the signatures of the $t\bar{t}H$ production include $e^+e^-\to t\bar{t}Z$, $t\bar{t}$, and $t\bar{t}g$ followed by $g\to b\bar{b}$. 
The cross sections for these background processes are plotted in Fig.\ref{fig:xsection} together with that of the signal. 
Notice the smallness of the contribution from the third diagram in Fig.\ref{fig:TTHdiagram}, which does not contain the top Yukawa coupling.
We can hence determine the top Yukawa coupling by just counting the number of signal events unless they are swamped by the background;
the signal cross section is only $\sim 0.5\,{\rm fb}$ with no beam polarization.

The production cross section for the $t\bar{t}Z$ background is $1.3\,{\rm fb}$\footnote{
This value is with QCD threshold enhancement similar to that expected for the signal process.
Without the correction the cross section is $0.7\,{\rm fb}$.
}
 with no beam polarization.  
It has four $b$-jets and two $W$s in the final state just like the signal, if the $Z$ boson decays into $b\bar{b}$ ($15\%$).
In this case the only difference that one can tell on an event-by-event basis lies in the invariant mass of the $b\bar{b}$ system, which should be consistent with $M_{H}$ for the signal and $M_{Z}$ for the background.
The $t\bar{t}$ production, on the other hand, has only two $b$-jets in the final state. 
If reconstructed correctly, it could not be the background.
Since the $t\bar{t}$ production cross section ($\sim 500\,{\rm fb}$) is much larger than that of the signal, however, a small fraction of mis-reconstruction or failure in $b$-tagging may lead to  significant background contamination.
The $t\bar{t}g$ production followed by $g\to b\bar{b}$ decay has the same signatures as the signal in terms of the number of $b$-jets and the number of $W$s.
As with the $t\bar{t}Z$ background the only difference is the invariant mass of the $b\bar{b}$ system.
Its production cross section is also of the same order, $0.7\,{\rm fb}$, as that of the $t\bar{t}Z$ background.

\section{Analysis Framework}
\begin{figure}[h]
       \begin{center}
         \includegraphics[scale=0.3,clip]{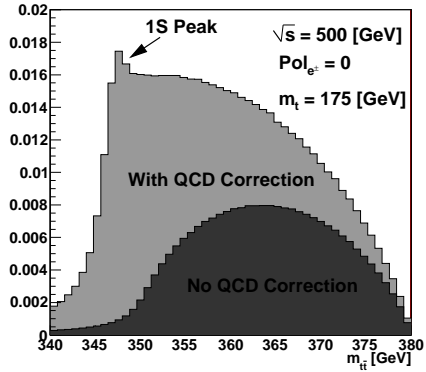}
        \end{center}
        \caption{Invariant mass distribution for the $t\bar{t}$ sub-system.}
        \label{fig:qcdcorrection}
   \end{figure}

For Monte Carlo simulations, we generated signal and background events by using an event generator package (physsim\cite{Ref:physsim}), which is based on full helicity amplitudes calculated with HELAS\cite{Ref:helas} including gauge boson decays, thereby correctly taking into account angular distributions of the decay products. 
The 4-momenta of the final-state quarks and leptons were passed to Pythia6.4\cite{Ref:pythia} for parton showering and hadronization. 
The resultant particles were then swum through a detector model (see Table \ref{Table:detparam} for detector parameters) defined in our fast Monte Carlo detector simulator (QuickSim\cite{Ref:quicksim}). 
In the event generations we used $\alpha(M_{Z}) = 1/128$, $\sin^2\theta_W=0.230$, $\alpha_{s} = 0.120$, 
$M_{W}=80.0\,{\rm GeV}$,
 $M_{Z}=91.18\,{\rm GeV}$,
 $M_{t}=175\,{\rm GeV}$, 
and $M_{H}=120\,{\rm GeV}$.
We have included the initial state radiation and beamstrahlung in the event generations. The unique point of this study is the inclusion of the QCD threshold enhancement to the $t\bar{t}$ system (see Fig.\ref{fig:qcdcorrection}) for the signal event generation, which plays an important role especially in a low energy experiment:
about a factor of $2$ enhancement at $\sqrt{s} = 500\,{\rm GeV}$. 

\begin{table}[h]
  \caption{Detector Parameters, where $p, p_{T}$ and $E$ are measured in units of GeV}
  \label{Table:detparam}
  \begin{center}
    \begin{tabular}{ccc}
      \hline
      Detector					&Performance														&Coverage\\
      \hline
      \hline
      Vertex detector			& $\sigma_b = 7.0 \oplus (20.0/p)/\sin^{3/2}\theta \mu m$			& $|\cos\theta| \le 0.90$	\\
      \hline
      Central drift chamber		& $\sigma_{P_{T}}/P_{T} = 1.1 \times 10^{-4}p_{T} \oplus 0.1\%$	& $|\cos\theta| \le 0.95$ \\
      \hline
      EM calorimeter			&$\sigma_{E}/E = 15\%/\sqrt{E} \oplus	1\%$						& $|\cos\theta| \le 0.90$	\\
      \hline
      Hadron calorimeter		&$\sigma_{E}/E = 40\%/\sqrt{E} \oplus	2\%$						& $|\cos\theta| \le 0.90$	\\
      \hline
    \end{tabular}
  \end{center}
\end{table}


\section{Event Selection}
\subsection{Definition of our signal (1-lepton+6-jet mode on $t\bar{t}H$)}
As explained in section \ref{sec:signature} we can classify the $t\bar{t}H$ signal events into the following three decay modes 
according to how the two $W$s from $t$ and  $\bar{t}$ decay: 
\begin{enumerate}
  \item 8-jet mode (45\%)
  \item 1-lepton + 6-jet mode (35\%)
  \item 2-lepton + 4-jet mode (7\%)
\end{enumerate}
where the lepton is required to be either $e^{\pm}$ or $\mu^{\pm}$ and the final-state $H$ to decay into the dominant $b\bar{b}$ state.
Notice that in all of these three modes we have four $b$-jets in the final states, which makes the separation of the $t\bar{t}$ background easier.
In this study we concentrate on the 1-lepton + 6-jet mode as our first step because the branching ratio is not so low and the number of jets is not so high. \\ 

\begin{figure}[h]
      \begin{minipage}{4.5cm}
        \begin{center}
          \includegraphics[scale=0.75,clip]{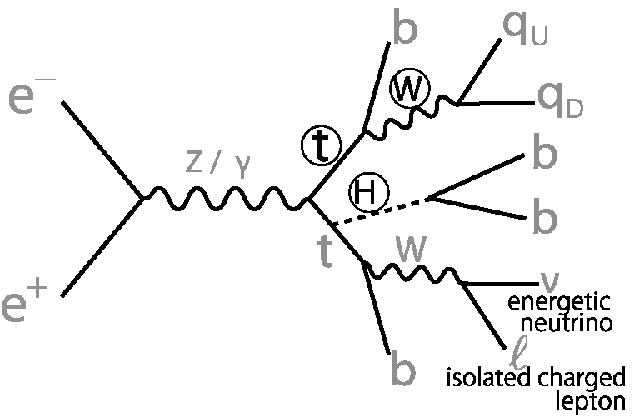}
          \caption{Schematic diagram defining our signal signatures}
          \label{fig:TTHdiagram2}
        \end{center}  
        \end{minipage}
        \hfill
        \begin{minipage}{9cm}
        \begin{center}
          \includegraphics[scale=0.43,clip]{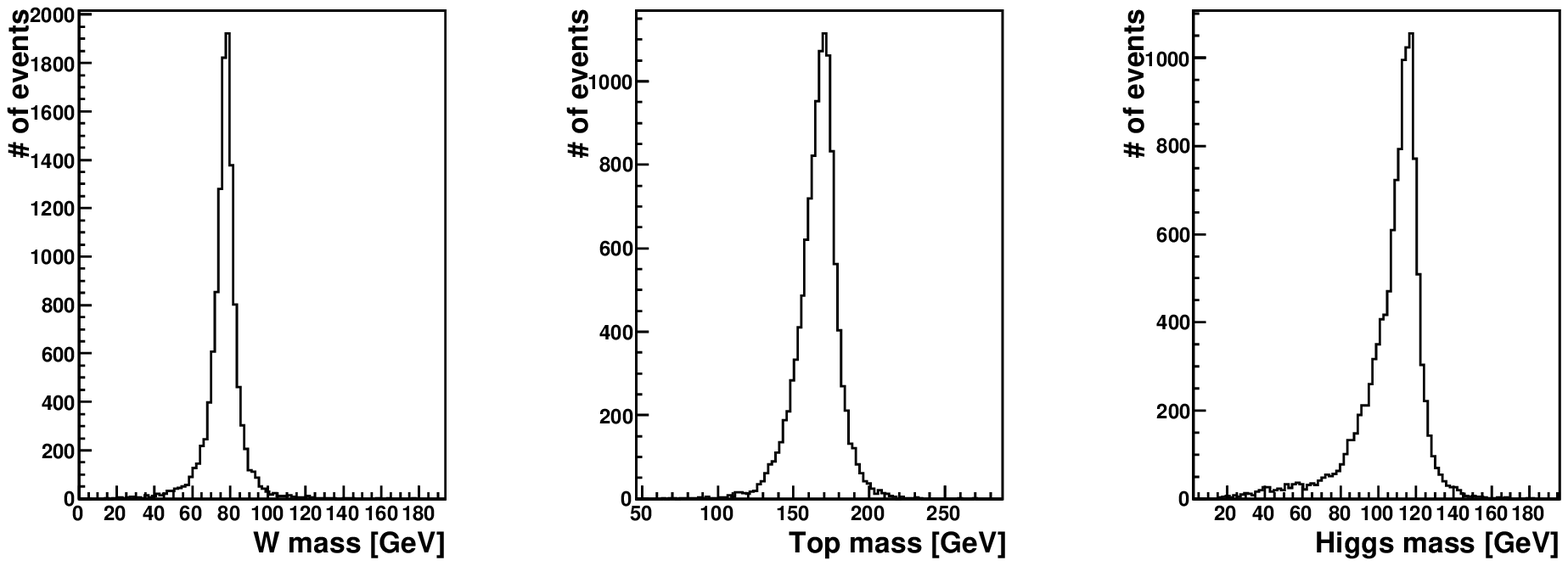}
          \caption{Invariant mass distributions for the hadronically decayed $W$, $t$, and $H$, which are reconstructed using generator information. }
          \label{fig:CheatMassSpec}
        \end{center}  
        \end{minipage}
      \end{figure}

As shown in Figs.\ref{fig:TTHdiagram2} and \ref{fig:CheatMassSpec} the signatures of our signal are 
\begin{itemize}
  \item an isolated energetic $e^{\pm}$ or $\mu^{\pm}$,
  \item six jets including four $b$-jets, two of which form a $H$ boson,
  \item the remaining two jets being consistent with a $W$ boson, and
  \item one of the two unused $b$-jets together with this $W$ candidate comprising a $t$ quark. 
\end{itemize}

In what follows we will elaborate selection cuts designed to single out these signatures.

  \subsubsection{Isolated lepton search}
\begin{figure}[h]         
     \begin{minipage}{6cm}
      \begin{center}
        \vspace{8mm}
        \includegraphics[scale=0.5,clip]{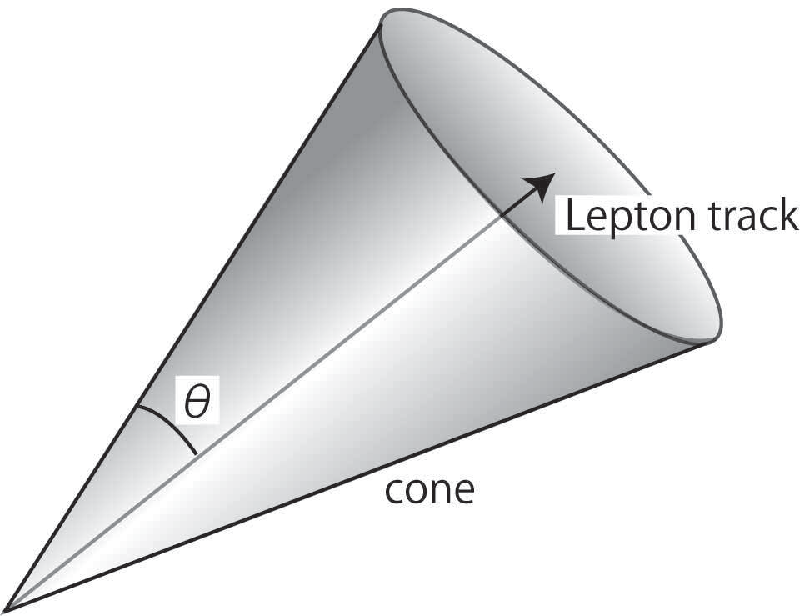}
        \vspace{5mm}  
        \caption{A cone around lepton track}
        \label{fig:cone}
       \end{center} 
     \end{minipage}
     \hfill
     \begin{minipage}{7.5cm}
        \begin{center}
          \includegraphics[scale=0.35,clip]{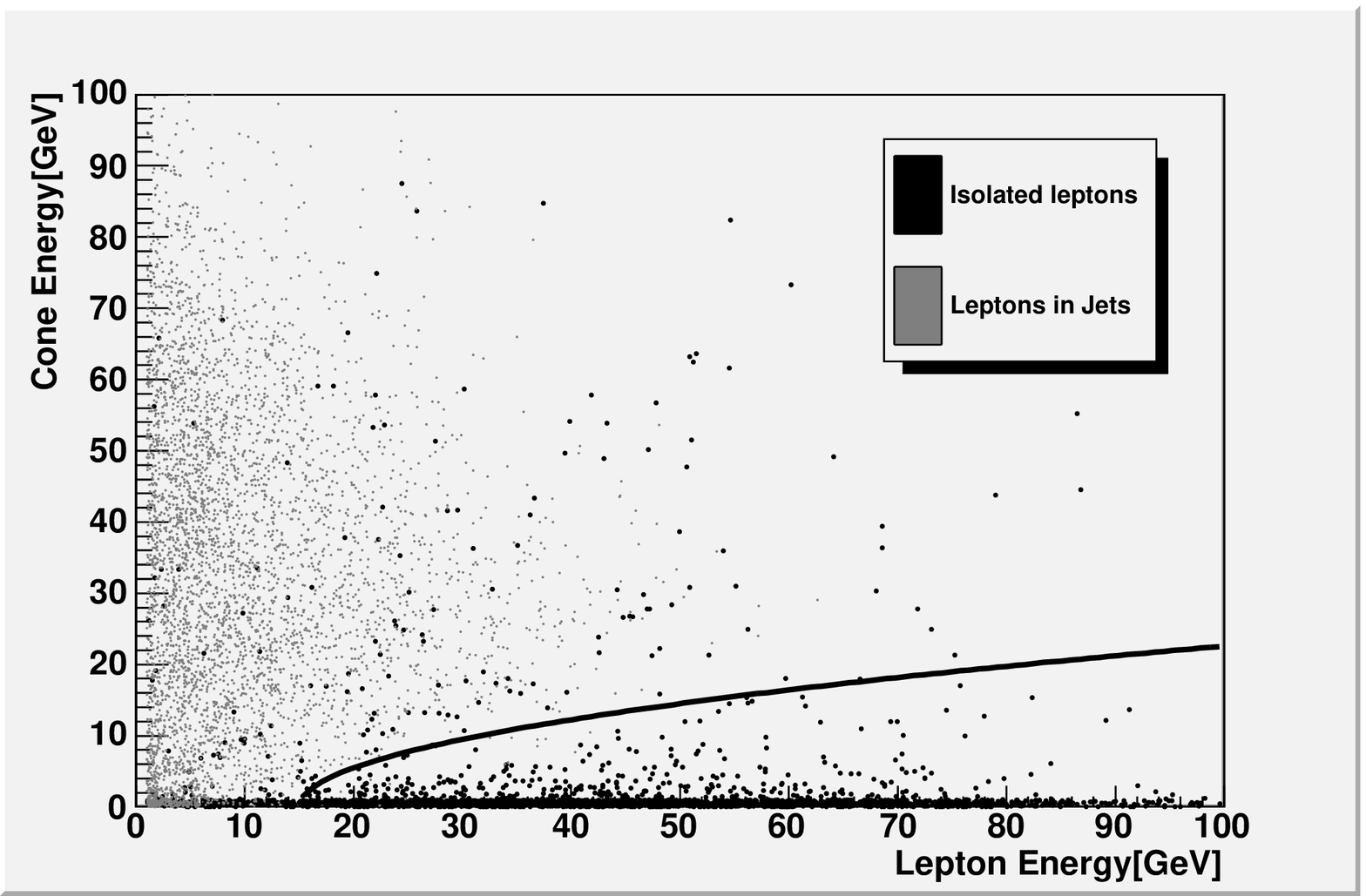}
          \caption{Cone energy distribution of isolated lepton: cut boundary $y=\sqrt{6(x-15)}$}
          \label{fig:isolepton}
        \end{center} 
       \end{minipage} 
      \end{figure}

  Our event selection starts with the search for a lepton coming from a $W \to l\nu$ decay. 
  Such a lepton from $W$ tends to be energetic and isolated from the other tracks.
  In order to find such an isolated lepton, we consider a cone around each lepton track (see Fig.\ref{fig:cone})
  and define the cone energy to be the sum of the energies of the other tracks in the cone.
Figure \ref{fig:isolepton} plots the cone energy against the lepton energy.
The energetic isolated leptons from $W$s have to have a high lepton energy and a low cone energy, 
hence populating the bottom edge region (black points), 
while leptons from heavy flavor jets are likely to be less energetic and have a higher cone energy (gray points).
The smooth curve in the figure is our cut to select energetic isolated leptons.

  \subsubsection{Forced 6-Jet clustering}
 After finding and eliminating an energetic isolated lepton, we perform jet clustering to make six jets.
 For the jet clustering we use a variable $Y$ defined by  
  \begin{eqnarray*}
  Y = \frac{M^2_{jet}}{E^2_{visible}}.
  \end{eqnarray*}
  We keep putting tracks together to form a jet while $Y < Y_{cut}$. 
  By adjusting the $Y_{cut}$ value, we can make arbitrary number of jets.
  We hence force the events to cluster into six jets by choosing an appropriate $Y_{cut}$ value on the event-by-event basis
  (forced 6-jet clustering).

  \subsubsection{$Y_{cut}$ cut}
  
\begin{figure}[h]
        \begin{center}
          \includegraphics[scale=0.5,clip]{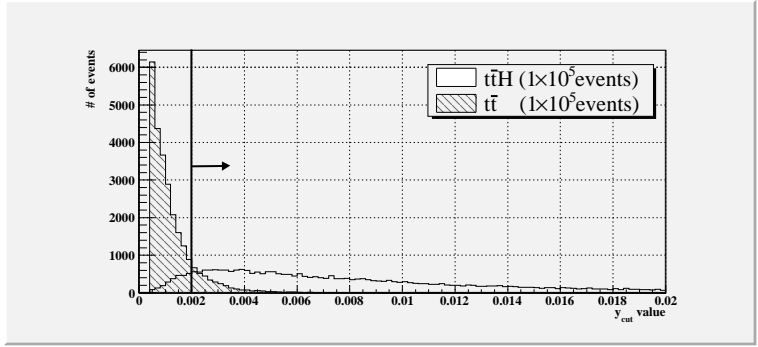}
          \caption{Ycut value distribution after isolated lepton finding}
          \label{fig:Ycut}
        \end{center}  
      \end{figure}

   The $Y_{cut}$ value for a $t\bar{t}$ background event to form six jets should be lower than the one for a signal  $t\bar{t}H\rightarrow t\bar{t}b\bar{b}$ event 
   because, after the energetic isolated lepton requirement, the $t\bar{t}$ event can hardly have more than four jets.
   The difference in the $Y_{cut}$ value distributions between $t\bar{t}H(H\rightarrow b\bar{b})$ and $t\bar{t}$ is shown in Fig.\ref{fig:Ycut}. 
   As seen in the figure, by cutting $Y_{cut}$ values at $0.002$ we can reduce the $t\bar{t}$ background effectively.
        
  \subsubsection{mass cut}
\begin{figure}[h]       
         \begin{center}
           \includegraphics[scale=0.6,clip]{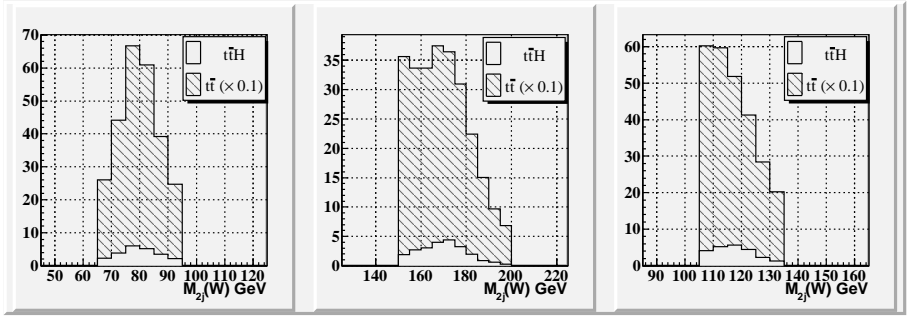}
           \caption{
                 Invariant mass distributions after the cut on $Y_{cut}$ values.
                 Black open histograms are for the signal and gray histograms are for the $t\bar{t}$ background. 
              }
           \label{fig:W_ycut}
         \end{center}  
      \end{figure}

  After performing the jet clustering, we try to identify which jet is coming from which parent parton. 
  We want to separate the correct combination from the other combinatorial background.
  Mass cut comes in handy to reduce the combinatorial background.
  Looping over all the 2-jet combinations we look for a pair having an invariant mass within the window of $\pm 15\,{\rm GeV}$ from the nominal $W$ 
  mass of $80.0\,{\rm GeV}$.
  From the remaining four jets we pick up one and attach it to the just found pair making a $W$ candidate to see if the resultant 3-jet system has an invariant mass
  within $\pm 25\,{\rm GeV}$ from the nominal $t$ mass of $175\,{\rm GeV}$.
  If it does we search for a pair from the three jets left over that is within the mass window of $\pm 15\,{\rm GeV}$ from the nominal $H$ 
  mass of $120\,{\rm GeV}$.  
  Since these mass cuts are rather loose there is a significant chance to have multiple combinations that pass them.
  For such a case we define a $\chi^2$ variable with 
  \begin{eqnarray}
  \nonumber
  \chi^2 &=& \left(\frac{M_{\mbox{2-jet}(W)}-M_{W}}{\sigma_{M_{W}}}\right)^2 
                         +\left(\frac{M_{\mbox{3-jet}(t/\bar{t})}-M_{t}}{\sigma_{M_{t}}}\right)^2
                        +\left(\frac{M_{\mbox{2-jet}(H)}-M_{H}}{\sigma_{M_{H}}}\right)^2  , 
  \end{eqnarray}
  and select the combination with the smallest $\chi^2$ value.
Fig.\ref{fig:W_ycut} shows the mass distributions for the best combinations.     
  Although $W$ and $t/\bar{t}$ peaks are present for both the signal and the $t\bar{t}$ background,
  a $H$ peak is seen only for the signal process.  
  The $H$ peak is, however, swamped in the $t\bar{t}$ background.

  \subsubsection{$b$-tagging by the $n$-sig. method}
  For the $t\bar{t}$ background rejection, $b$-tagging is very powerful since the signal $t\bar{t}H(H\rightarrow b\bar{b})$ process has four $b$-jets, 
  while the $t\bar{t}$ background process has only two $b$-jets.
  For $b$-tagging we use the so called $n$-sig. method descrived as follows.\\
  
\begin{figure}[h]
       \vspace{5mm}
        \begin{center}
          \includegraphics[scale=0.6,clip]{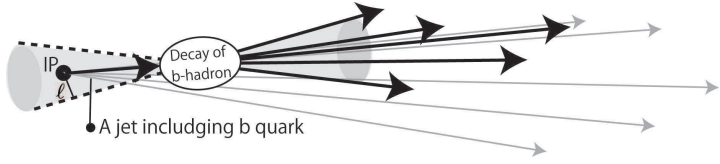}
          \vspace{2mm}
          \caption{$n$-sig. method}
          \label{fig:btag}
        \end{center}  
      \end{figure}

  Figure \ref{fig:btag} sketches a jet from the interaction point (IP), which includes a $b$-hadron.
  The $b$-hadron decays at distance from the IP due to its long-life.
  It makes the $b$-jet to have some tracks which are away from the IP.
  When the distance ($\ell$) between the IP and a track is larger than a given value ($m \sigma_{\ell}$), the track is defined as an off-vertex track.
  A jet is recognized as a $b$-jet if the number of such significantly off-vertex tracks exceeds a certain cut value ($n$).
  In this analysis, we define tight $b$-tagging with a tagging condition: $(m,n)=(3.0,2)$
  and loose $b$-tagging with $(m,n)=(2.0,2)$, 
  and require all of the four $b$-jet candidates have to satisfy the loose $b$-tagging condition and there has to be at least one tight $b$-tagged jet from each of the $H$ and $t/\bar{t}$ candidates.
  \clearpage 
  
\begin{figure}[h]       
         \begin{center}
           \includegraphics[scale=0.6,clip]{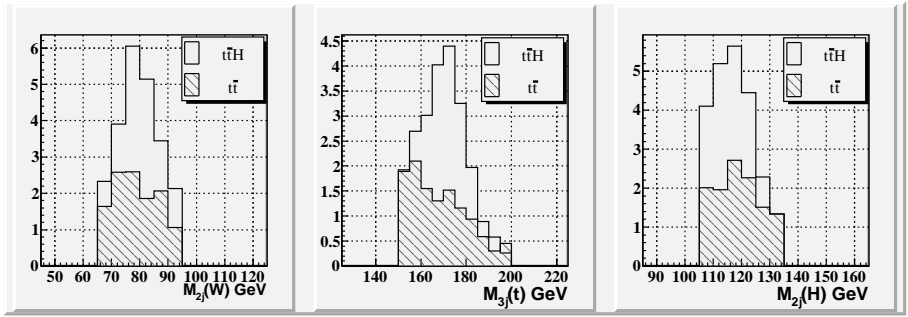}
           \caption{Invariant mass distribution after using both Y cut and b-tagging}
           \label{fig:W_ycut_btag}
         \end{center}  
      \end{figure}

The mass distributions after the $b$-tagging are shown in Fig.\ref{fig:W_ycut_btag}.
We can see that the $t\bar{t}$ background has been suppressed effectively.
As mentioned above the $t\bar{t}Z$ and $t\bar{t}g$ ($g\to b\bar{b}$) background events have similar signatures as a signal and can be separated only with the invariant mass of the $H$ candidate.
In the next section we summarize the results of our event selection including these remaining background processes.
        



\section{Results}
In order to estimate the feasibility of measuring the top Yukawa coupling we need to specify the beam polarization and the integrated luminosity.
In this study we assume an integrated luminosity of $1\,{\rm ab}^{-1}$.
As for the beam polarization, it is worth noting that only the left-right or right-left combination contributes to the signal and background cross sections because of the $\gamma^\mu$ coupling of the beam particles to the vector bosons ($\gamma/Z$) in the intermediate states.
It is hence sufficient to know the cross sections for the beam polarization combinations: $(e^-,e^+) = (-1,+1) ,(+1,-1)$.
Table\ref{tab:Xsection} shows these cross sections.

\begin{table}[h]
  \caption{Cross sections at $\sqrt{s}=500\,{\rm GeV}$.  $t\bar{t}H$ and $t\bar{t}Z$ are with QCD threshold enhancement. 
  (-1,+1)/(+1,-1) corresponds to $(e^{-}_{L},e^{+}_{R})/(e^{-}_{R},e^{+}_{L})$, respectively.}
  \label{tab:Xsection}
  \begin{center}
    \begin{tabular}{ccc}
      \hline
      Beam Polarization				&  (-1,+1)		& (+1,-1)\\
      \hline
      \hline
      $t\bar{t}H$						& 1.24 [fb]		& 0.540 [fb]		\\
      \hline
      $t\bar{t}Z$						&2.18 [fb]		& 0.712 [fb]		\\
      \hline
      $t\bar{t}$						&720. [fb]		&309. [fb]		\\
      \hline
      $t\bar{t}g$ ($g\to b\bar{b}$)	&1.93 [fb]		&0.859 [fb]\\
      \hline
    \end{tabular}
  \end{center}
\end{table}

For both of the beam polarization combinations: $(-1,+1)$ and $(+1,-1)$,
 we have generated $50{\rm k}$ events each for the $t\bar{t}H$, $t\bar{t}Z$, and $t\bar{t}g\,(g\to b\bar{b})$ processes, 
 and $5{\rm M}$ events for the $t\bar{t}$ background.
We performed the event selection described in the previous section and tabulated the results normalized to an integrated luminosity $1\,{\rm ab}^{-1}$ in Table\,\ref{tab:allpol} assuming the cross section shown in Table\,\ref{tab:Xsection}. 
\begin{table}[h]
  \caption{Cut Statistics (normalized to 1$\,\rm{ab}^{-1}$)}
  \label{tab:allpol}
  {\footnotesize
  \begin{center}
    \begin{tabular}{c||cccc|cccc}
      \hline
      Beam Polarization	&\multicolumn{4}{c|}{(0.0,0.0)} &\multicolumn{4}{c}{(-0.8,+0.3)} \\
     Processes							& $t\bar{t}H$	& $t\bar{t}Z$	& $t\bar{t}$	&$t\bar{t}g\,(b\bar{b})$	& $t\bar{t}H$	& $t\bar{t}Z$	& $t\bar{t}$	&$t\bar{t}g\,(b\bar{b})$\\
      \hline
      \hline
      No Cut								& 449.0		& 1340.0		& 514040.5	&697.5						& 759.0		& 2407			& 863500.4	&1159.6\\
      \hline
       $N_{iso.lep}$=1						& 159.4		& 435.9		& 209718.4	&242.2						& 269.4		& 783.0		& 303879.0	&397.7\\
      \hline
      $Y_{cut}\,\mbox{(6 jets)} > 0.002$	& 139.2		& 307.8		& 22851.3		&152.5						& 235.4		& 552.9		& 38477.2		&249.6\\
      \hline
      btag \& mass cut						& 23.0      		& 12.2			& 11.9			&6.9						& 38.9     		& 21.8			& 19.7			&11.3\\
      \hline
    \end{tabular}
  \end{center}
  }
\end{table}

\begin{figure}[h]       
         \begin{center}
           \includegraphics[scale=0.6,clip]{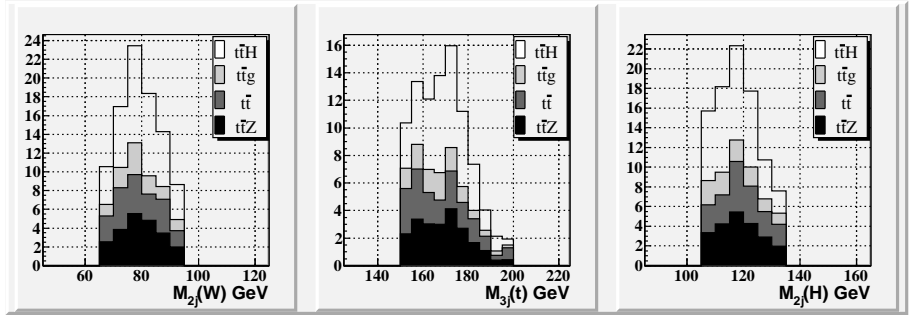}
           \caption{
               Mass distributions (cumulative) for the final selected sample for the beam polarization combination: $(-0.8,+0.3)$.
           }
           \label{fig:massfinal-08+03}
         \end{center}  
      \end{figure}

The corresponding distributions for the reconstructed $W$, $t/\bar{t}$, and $H$ candidates are shown in Fig.\ref{fig:massfinal-08+03} for the beam polarization combination: $(-0.8,+0.3)$.
We can see a clear evidence of signal events over the background in each of the three mass distributions.

In the case of no beam polarization $23.0$ signal events are left with $31.0$ background events total. 
On the other hand we have $38.9$ signal events with $52.8$ background events total at the end of the event selection.
The signal significance is $4.1\,\sigma$ for the polarization combination: $(0,0)$ and $5.4\,\sigma$ for the polarization combination: $(-0.8,+0.3)$.
Since the number of the signal events is proportional to the square of the top Yukawa coupling ($g_{Y}$), 
we can easily translate these numbers to its expected precisions: 
$\Delta g_{Y}/g_{Y} = \pm 0.12$ and $\pm 0.093$ for the beam polarization polarization combinations: $(0,0)$ and $(-0.8,+0.3)$, respectively.

\section{Summary and Conclusion}
We have performed a feasibility study of measuring the top Yukawa coupling at $\sqrt{s}=500\,{\rm GeV}$, taking advantage of the QCD threshold enhancement to the $t\bar{t}$ sub-system.
For this study we have implemented the threshold enhancement in the $t\bar{t}H$ and $t\bar{t}Z$ event generators in the physsim package.
It is found that for an integrated luminosity of $1\,{\rm ab}^{-1}$ we can observe the $t\bar{t}H$ process with a significance of $4.1\,\sigma$ without beam polarization,
and $5.4\,\sigma$ with the beam polarization combination: $(e^-,e^+)=(-0.8,+0.3)$. 
These numbers show that we can measure the top Yukawa coupling to an accuracy of about $10\%$ at $\sqrt{s}=500\,{\rm GeV}$, which is the energy already available in the first stage of the ILC.

\section{Acknowledgments}
The authors would like to thank all the members of the ILC physics subgroup
\cite{softg} for useful discussions. 
Among them, A.\,Ishikawa, Y.\,Sumino, and Y.\,Kiyo deserve special mention for their contributions to the implementation of the QCD threshold correction.
They are also grateful to T.\,Tanabe for providing us with the $t\bar{t}g\,(g\to b\bar{b})$ generator.
This study is supported in part by the Creative Scientific Research Grant
No. 18GS0202 of the Japan Society for Promotion of Science and the JSPS Core University Program. 

\begin{footnotesize}

%
\end{footnotesize}